\documentstyle[12pt]{article}
\begin{document}
\begin{center}
{\large Lattice animals and the Percolation model under rotational constraint}
\vskip 1cm
                        Indrani Bose \vskip .5cm
                        Department of Physics\\
                        Bose Institute\\
                        93/1, A. P. C. Road\\
                        Calcutta-700 009, India.
\end{center}
\begin{abstract}
The effect of rotational constraint on the properties of lattice models like
the self-avoiding walk, lattice animals and percolation is discussed. The
results obtained so far, using a variety of exact and approximate techniques,
are described. Examples of the rotational constraint in real systems are also
given.
\end{abstract}
\section*{I. Introduction}
Lattice models describe a wide variety of physical, chemical and biological systems.
The systems either have an underlying lattice structure or are studied on a lattice
for convenience. In the latter case, only those properties of the system are
studied which are independent of the microscopic structural details of the
system. In this article, we will discuss three lattice models: the self-avoiding
walk (SAW), lattice animal (LA), and percolation. We will mainly discuss the effect
of rotational constraint on the various configurational and connectivity
properties of the models.

A SAW is a walk on a lattice with the constraint that a site once visited 
cannot be visited again. The SAW is a model of linear polymers in dilute 
solution \cite{Gennes}. A linear polymer can be pictured as a string of smaller
chemical units called monomers. There is a one-to-one correspondence between each 
possible configuration of the polymer and a particular SAW on the lattice.
The self-avoidance criterion is due to the excluded-volume effect of
polymers which implies that no two monomers can occupy the same physical
space. 

LAs are clusters of connected sites (site LA) or bonds (bond LA) embedded
in a lattice. The LAs represent random clusters which can occur in a variety 
of systems. In terms of polymers, one example is branched polymers in
dilute solution. The percolation model is a model of disorder and has, in 
general, two versions, site and bond. In the site (bond) percolation
model, the probability that a site (bond) of the lattice is occupied is
p. For p less than a critical value $p_c$, called the percolation threshold, 
no extended network of sites (bonds) spans the lattice. For p > $p_c$,
such a network exists with probability one. The percolation threshold
$p_c$ has a unique value for an infinitely large lattice. The value of
$p_c$ is different for different lattices. The percolation transition at 
p = $p_c$ is a phase transition, often called a geometrical phase 
transition. In one phase ( p > $p_c$ ), an infinite cluster exists and 
in the other phase ( p < $p_c$ ), there is no infinite cluster. In the
vicinity of p = $p_c$, the percolation model exhibits critical 
phenomena analogous to those exhibited by a second-order thermodynamic
phase transition. The percolation model has extensive applications in
physics, chemistry and biology \cite{Stanley1,Stauffer1,Bunde,Stauffer2}.

We now show the connection between LAs and the percolation problem. We
consider the site cluster problem. The generating function desfribing an 
ensemble of clusters of all sizes is written as
\begin{equation}
G\,(\,x,y\,)\,=\,\sum_{s,t}\,g_{s,t}\,x^s\,y^t
\end{equation}
$g_{s,t}$ is the number of distinct LAs of size s and t perimeter sites, 
each animal being weighted by a factor $x^s\,y^t$. The size s of the 
animal is equal to the number of sites in the cluster. In the percolation
problem, each cluster site is occupied with probability p and a perimeter
site being an unoccupied site has the probability 1-p of being so. Thus
x = p and y = 1-p in Eq.(1).The corresponding generating function is
\begin{eqnarray}
G\,(\,p,q\,)\,&=&\,\sum_{s,t}\,g_{s,t}\,p^s\,q^t\nonumber\\
              &=&\,\sum_s\,D_s(q)\,p^s  
\end{eqnarray}
where q = 1-p and $D_s(q)\,=\,\sum_t\,g_{s,t}\,q^t$.The $D_s(q)$ are known as
perimeter polynomials. For LAs, clusters of the same size have equal weight
irrespective of the number of perimeter sites, so y = 1 in Eq.(1) and the
generating function becomes 
\begin{equation}
G\,(\,x,1\,)\,=\,\sum_s\,g_s\,x^s
\end{equation}
where $g_s$ = $\,\sum_t\,g_{s,t}$ is the total number of clusters of size s.
Eq.(2) shows that percolation and LA models are related through the 
perimeter polynomials construction of which requires knowledge of LA 
numbers.Both generating functions (2) and (3) become singular as 
$p\,\rightarrow\,p_c$ and $x\,\rightarrow\,x_c$, respectively, hence 
$p_c$ and $x_c$ are identified as critical points of the respective
ensembles of clusters. Below the critical point small clusters are 
predominant in number, above the critical point the reverse situation 
is true. At the critical point, clusters of all sizes exist, a familiar 
characteristic of critical phenomena. In fact, the transition is a second-order 
transition and the generating function in each case plays the role of
free energy in thermal critical phenomena. Analogous to thermodynamic
quantities, cluster-related quantities can be defined for the geometric
models. These show singular behaviour as the critical point is being reached.
The singularity is of the power-law type with an associated critical
exponent. For the SAW model, a generating function similar to Eq.(3)
can be defined with s being the number of steps in the walk. For generating 
functions of the type (3), it can be shown \cite{Family1} that every critical
exponent characterising a singularity as $x\,\rightarrow\,x_c$, say, is
uniquely related to a critical exponent along the path
$s\,\rightarrow \,\infty$, i.e., s $\sim\,{\mid\,x - x_c\mid\,}^{-1}$. In the
asymptotic limit of s going to infinity, various walk or cluster properties
scale with size. We will now describe two such relations for site LAs.
Equivalent relations can be written down for SAWs. The first relation shows 
that the total animal (cluster) number $g_s$ scales as
\begin{equation}
g_s\,\,\sim_{s\rightarrow\infty}\, \lambda^s\,s^{-\theta}
\end{equation}
where $\lambda$ is a constant for a particular lattice and is known as
the `growth parameter' because asymptotically
$g_s\,/g_{s-1}\,\rightarrow\,\lambda$. The parameter $\theta$ (critical
exponent) is `universal' in the sense that it has the same value for all 
lattices in dimension d, the value changes only if d changes.The second
quantity we are interested in is the average radius of gyration defined
as $R_s\,=\,(<\,\sum_{i=1}^{s}\,r_{i}^{2}\,/s\,>)^{1/2}$ where $r_i$ is the 
distance of a cluster site i from the centre of mass of the cluster and
<....> denotes average over all animals. As ${s\,\rightarrow\,\infty}$,
$R_s$ scales with s as
\begin{equation}
R_s\,\sim\,s^\nu
\end{equation}
where $\nu$ is another critical exponent. In connection with relations
like (4) and (5), it has been shown\cite{Stauffer2} that LAs and large
clusters below the percolation threshold have identical scaling behaviour.
For the SAW model, $g_s$ is the number of distinct SAWs with s steps and
$R_s$ is the root-mean-square end-to-end displacement for walks with s
steps.

The scaling relations (4) and (5) have been well-studied for both 
undirected and directed models. In the case of undirected SAW and LA, 
there is no constraint on connectivity. In the case of directed SAW 
and LA, there is a global directionality constraint. To give an example,
starting from the origin and while occupying new sites, the animal can
grow only upwards and towards the right. The undirected and directed lattice
models are known to belong to different universality classes, i.e., the 
exponents $\theta$ and $\nu$ have different values. A new type of constraint
\cite{Privman,Li,Bose1,Bose2} that has been proposed is the rotational
constraint. In the presence of this constraint, a SAW or a LA can grow
either in the same direction as it has been growing in or in a specific
rotational direction, say, clockwise. Similar constraint can also be
defined for the percolation model. In Section II, the SAW model under
rotational constraint, termed the spiral SAW (SSAW), is described. In
Section III, spiral LAs are considered. A brief description is also
given of directed compact site LAs for which the generating function
can be determined in an exact manner. In Section IV, the spiral
percolation model is discussed. Section V contains some concluding
remarks.

\section*{II. Spiral SAW}
The model was first considered by Privman\cite{Privman}. Let us consider 
the SSAW problem on a square lattice. Each step of the walk may either point
in the same direction as the preceding one or in a direction rotated by
$\,+\,\pi/2$ w.r.t. it. The global behaviour arising from the two
constraints of self-avoidance and the bond angle restrictions is quite
novel. In general, a SSAW consists of an outward spiralling part and
an inward spiralling part (Fig.1). Each SAW is composed of horizontal
and vertical segments. Let $s_n$ be the number of n-step SSAWs. In the
asymptotic limit, exact, analytical expressions for $s_n$ have been
determined \cite{Blote, Guttmann1,Guttmann2,Guttmann2,Lin} in the cases
of both the square and triangular lattices. We describe briefly the 
results obtained in Refs. [12-13] for the square lattice. 

Let the first step of the walk be vertically upwards. Each SSAW can be
decomposed into a concatenation of two walks of type I and C by either a 
vertical line (Fig.2(a)), a horizontal line (Fig.2(b)) or in some cases
both (Fig.2(c)).$I_n$ denotes the set of n-step SSAWs whose last and third
last segments are of equal length, $i_n$ is the number of such walks.
$C_n$ is the set of n-step SSAWs whose last segment is at least one
greater than the length of the third last segment and $c_n$ the number
of such walks. From Figs. 2(a) - 2(c), the section of the spiral from
O to A belongs to $I_k$ and that from A to N belongs to $C_{n-k}$. The
walks of the type shown in Figs. 2(a) - 2(b) can be uniquely decomposed
in this way. Thus
\begin{equation}
S_n\,=\,\sum_{k=0}^{n-1}\,i_k\,c_{n-k}\,-\,\sum_{k=1}^{n-1}\,i_k\,i_{n-k}
\end{equation}
with $i_0$ = 1, $c_0$ =0, $c_1$ =1. The second sum corrects for a 
double-counting of walks of the type 2(c) by the first sum. A geometrical 
construction\cite{Guttmann1} transforms walks of the type 2(c) into
two connected walks from the set I. The walk
in the set $C_{n+1}$ can be obtained by adding a step in a particular
direction to each member of $C_n\,U\,I_n$. Thus
\begin{equation}
c_{n+1}\,=\,c_n\,+\,i_n
\end{equation}
Now $c_n$ is equal to the number of partitions of the integer n. An
example is given in Fig.(3). Fig.3(a) represents a particular partition
of the number 12. A line at $45^{\circ}$ divides the set of points into two 
classes, representing vertical and horizontal segments. There are two 
horizontal segments of length 2 and 3 and two vertical segments of length
3 and 4. The resultant SSAW is shown in Fig.3(b). Each partition of the
integer 12 can be represented by a pattern of points and each such pattern
can be uniquely represented by a SSAW. All the spiral walks in the set 
$C_n$ can be enumerated in this way.

Asymptotic expressions (n $\rightarrow\,\infty$ ) for the number of 
partitions of integers are known from the classic work of Hardy and 
Ramanujan\cite{Hardy}. One obtains
\begin{equation}
c_n\,=\,\frac{1}{4\sqrt{3}\,n}\,e^{\pi\,(\,\frac{2\,n}{3}\,)^{\frac{1}{2}}}\,[1\,+\,\frac{c}{\sqrt{n}}\,+\,O\left(\frac{1}{n}\right)]
\end{equation}
where c is a constant which can be determined.
From (7),
\begin{equation}
i_n\,=\,\frac{\pi}{12\,\sqrt{2}\,n^{\frac{3}{2}}}\,e^{\,\pi\,(\,\frac{2\,n}{3}\,)^{\frac{1}{2}}}\,\big[1\,+\,O\left(\frac{1}{\sqrt{n}}\right)\big]
\end{equation}
Substituting (8) and (9) into (6), one finds that the largest terms 
in the sum are in the vicinity of k = n/2. Expanding around the 
maximum and replacing the sum by an integral, one finally obtains
\begin{equation}
s_n\,=\,\left(\frac{\pi}{4 \times \,3^{\frac{5}{4}}}\right)\,n^{\frac{-7}{4}}\,exp\big[2\,\pi\,\left(\frac{n}{3}\right)^{\frac{1}{2}}\big]\,\big[1\,+\,O\left(\frac{1}{\sqrt{n}}\right)\big]
\end{equation}
Joyce\cite{Joyce} has obtained the complete asymptotic expansion for 
$s_n$. Lin\cite{Lin} has obtained an expression for $s_n$ in the case of
the triangular lattice. By comparing the expression (10) with that (Eq.(4))
valid for undirected and directed SAWs, one finds that the rotational
constraint produces a new scaling form for the number of SAWs. Bl\"{o}te
and Hilhorst\cite{Blote} have also obtained an asymptotic form for the 
root mean square end-to-end distance of SSAWs. Again, the rotational
constraint has a non-trivial effect on the scaling form. 

\section*{III. Spiral site lattice animals}
Two versions of the spiral LA have been studied : `rooted', i.e., the 
origin from which the animal is grown is kept fixed and `unrooted', i.e., the
origin is not specified. In the latter case, two animals differing only
in the choice of the origin are considered identical. The number of distinct
rooted animals of a certain size is much greater than the number of 
unrooted animals of the same size. Bose and Ray\cite{Bose1} have defined
unrooted spiral LAs in d = 2 to be a subset of undirected and unrooted 
LAs with the proviso that there is a site of the cluster,namely, the
origin to which every other site of the cluster is connected through at least
one spiral path. In a spiral path, connection is either in the forward 
direction or in a specific rotational direction,say, clockwise. An example
of a spiral site LA is shown in Fig.4. In dimension d > 2, a possible
definition of the spiral constraint is: the projection of the path joining
any site of the animal to the origin (the first site to be occupied), on
any specified plane, say the xy plane, should have no anticlockwise turns.
Other definitions of the spiral constraint have also been suggested\cite{Bose2}.

A variety of techniques can be used to study the configurational properties
of LAs.For spiral LAs, the method of exact enumeration of clusters upto a
certain size has been used to obtain the values of $\lambda$ and the 
exponents $\theta$ and $\nu$ in the scaling relations (4) and (5). Unlike 
in the case of SSAWs, the scaling relations (4) and (5) appear to give an 
adequate description of the configurational properties of spiral LAs. The
exact enumeration of animals of a certain size s is done on the computer using
the well-known Martin's algorithm\cite{Martin}. This algorithm generates
undirected LAs and then each animal is checked for spiral connection.
Knowledge of LA numbers of various sizes enables one to calculate the
ratios $g_{s+1}/\,g_s$ for various values of s. From Eq.(4), in the large
s limit, the plot of the ratio $g_{s+1}\,/g_s$  versus 1/s is that of a 
straight line. From the intercept and slope of the straight line, the 
growth parameter $\lambda$ and the exponent $\theta$ can be determined.
For spiral site LAs on the square lattice, exact enumeration data is
available for size upto s = 15\cite{Santra1}. The values of $\lambda$
and $\theta$ are determined as $\lambda\,=\,3.002\,\pm\,0.020$ and
$\theta\,=\,0.69\,\pm\,0.10$. The radius-of-gyration exponent 
$\nu\,=\,0.50\,\pm\,0.02$, i.e., the LAs are nearly-compact. These values
are different from the corresponding values in the cases of undirected
and directed LAs. Thus spiral LAs belong to a new universality class.

For the undirected and directed LAs, it is well-known that animals
and trees (animals without loops) belong to the same universality class, i.e.,
loops have no effect on cluster statistics. This fact has been arrived at
through detailed studies using techniques like both field-theoretic and 
position space RG, series-expansion and Monte Carlo enumeration\cite{Lubensky,
Family2,Daoud,Duarte}. For spiral trees, the values of $\lambda_0$, 
$\theta_0$ and $\nu_0$ have been obtained as\cite{Santra1} (from exact 
enumeration data upto s = 15) $\lambda_0\,=\,2.123\,\pm\,0.004$, 
$\theta_0\,=\,-1.318\,\pm\,0.020$ and $\nu_0\,=\,0.67\,\pm\,0.02$. These
values are different from the corresponding values for spiral LAs. Thus the 
spiral LAs and trees belong to different universality classes, i.e., 
loops have a non-trivial effect on spiral animal statistics. It has as yet
not been possible to rigorously explain this fact but a special feature of
spiral clusters is worth mentioning. For both ordinary and directed
animals, an animal with loops becomes a tree as soon as a minimal number of 
bonds ( for bond LAs) or sites (for site LAs) are removed but this is not
always the case for a spiral animal. A tree obtained from a spiral animal
with loops need not be a spiral tree. Also, in order to grow in the 
forward direction, a spiral animal should contain many loops. Without
the loops, spiral connectivity is not possible in many cases. How these
facts affect asymptotic cluster charateristics is still not understood.

An interesting result associated with the LA problem is that of dimensional
reduction. Parisi and Sourlas\cite{Parisi} have shown that the asymptotic
properties of LAs in d dimensions are the same as the critical properties
of the Lee-Yang edge singularity (LYES) of the Ising model in d - 2 
dimensions. For the directed LA problem, the dimensional reduction is
d - 1\cite{Stanley2}. The LYES problem has to do with the fact that 
ferromagnetic Ising models, besides exhibiting singular behaviour at the
critical point T = $T_c$, in the presence of zero external magnetic field, show
singular behaviour also at a critical value of imaginary magnetic field
$H\,=\,H_0\,(T)$ and for T > $T_c$. The magnetiation has a branch point
singularity of the form
\begin{equation}
M\,-\,M_0\,\sim\,(\,H\,-\,H_0\,)^{\sigma}
\end{equation}
where $M_0$ is the magnetization at $H_0$. Consequently, the free energy F of
the (d - 2)-dimensional Ising model has the same singular structure as that of 
the generating function of the d-dimensional undirected LA problem. From
Eqs.(3) and (4), it follows that the LA generating function G(x) has a 
singularity of the form $(x\,-\,x_c)^{\theta-1}$ where $x_c\,=\,1/\lambda$.
Since magnetization is the first derivative of the free energy F w.r.t. the 
magnetic field H, F has a power-law singularity of the type (11) with the
exponent $\sigma\,+\,1$. From the equivalence of the free energy and the
generating function in appropriate dimensions, one obtains
\begin{equation}
\theta(d)\,=\,\sigma(d-2)\,+\,2
\end{equation}
The Josephson scaling law (the hyperscaling relation) connects the singularity
of the free energy F with the singularity of the correlation length $\xi$, 
$(F\sim\,\xi^{-d})$. One thus obtains
\begin{equation}
\nu(d)\,=\,[\,\sigma(d-2)\,+\,]/(d-2)
\end{equation}
Each undirected and unrooted LA of N sites gives rise to N distinct rooted
site animals. Thus the animal number exponents, in the two cases, differ by
1. Thus, from (12) and (13), one gets 
\begin{equation}
\theta\,=\,(d-2)\,\nu
\end{equation}
for rooted undirected animals. Similarly, one can show that
\begin{equation}
\theta\,=\,(d-1)\,\nu_{\perp}
\end{equation}
for rooted directed animals. The exponent $\nu_{\perp}$ is associated with
the radius of gyration of the directed animals perpendicular to the
preferred direction of growth. We have seen that the LA exponents $\theta$,
$\nu$ are related to the LYES exponent $\sigma$. The value of $\sigma$ can
be calculated exactly for d = 0 (point particle) and d = 1.Thus from
Eqs.(12) and (13), the exponents $\theta$, $\nu$ can also be determined
exactly in appropriate higher dimensions (upto d = 3). The directed LA
problem is exactly solvable in both 2d and 3d\cite{Dhar1}.So exploiting
the fact that directed animal exponents in d dimensions are related to the 
ordinary animal exponents in (d + 1) dimensions and to the LYES exponents in (d-1)
dimensions, $\theta$, $\nu$ are exactly known for ordinary animals in d = 4
and $\sigma$, the LY exponent, in d = 2. The results for $\sigma(d)$, 
$\theta(d)$, $\nu(d)$ are: $\sigma(0)\,=\,-1,\sigma(1)\,=\,-1/2,
\,\sigma(2)\,=\,-1/6$, for ordinary animals $\theta(2)\,=\,1, \theta(3)\,=\,
3/2,\,\theta(4)\,=\,11/6,\,\nu(3)\,=\,1/2,\,\nu(4)\,=\,5/12$ and for
directed animals $\theta(2)\,=\,1/2,\,\theta(3)\,=\,5/6,\,\nu_{\perp}(2)\,
=\,1/2,\,\nu_{\perp}(3)\,=\,5/12$.

To study dimensional reduction in the spiral animal problem, Bose et al\cite
{Bose2} have considered only spiral trees. Both exact enumeration and MC
simulation have been performed. The spiral trees considered are rooted and the
rotational constraint is operative only in the xy plane. The radius of gyration has
two components $R_{s,pl}$ in the xy plane and $R_{s,\perp}$ in the direction
perpendicular to the xy plane. The numerical data obtained are consistent
with the following results:
\begin{eqnarray}
\theta_0\,&=&\,(d-4)\,\nu_{0,pl}\,\,\, (d=2)\nonumber\\
\theta_0\,&=&\,(d-4)\,\nu_{0,\perp}\,\,\,( d > 2)     
\end{eqnarray}
Eq.(16) shows that a dimensional reduction by 4 occurs in the spiral animal
problem. The dimensional reduction by the factor two, in the case of 
ordinary animals, occurs due to a hidden supersymmetry\cite{Parisi,Sourlas}.
The first few applications of supersymmetry in condensed matter and statistical 
physics include the following problems: the random field model, the linear
polymers (SAWs on a lattice), the branched polymers (LAs) and electron
localization. These models can be suitably recast into forms which
manifest supersymmetry. Introduction of an effective supersymmetry simplifies 
technically complicated calculations.Also, as in the cases of branched polymers
and random field models, which are closely related, the supersymmetry can be
shown to lead to an effective reduction in the number of degrees of freedom. 
For the directed animal problem, the dimensional reduction by 1 occurs because 
the disorder or cellular automaton condition is satisfied. Field theories have
been formulated for both ordinary and directed animal problems, specially, 
with the view to understand the mechanism of dimensional reduction. For
spiral LAs, a complete theory is yet to be worked out. A simple example of
reduction of effective dimensionality due to rotational constraint, is
provided by the quantum mechanical motion of a charged particle in a 
uniform magnetic field. If the field is perpendicular to the xy plane, then 
the degrees of freedom in this plane get quantized. The resulting discrete 
spectrum has no contribution to the density of states of low energy
excitations. This leads to a reduction by 2 in the effective dimensionality.
One can argue, in analogy, that for the spiral animal problem also, the rotational
constraint is responsible for a dimensional reduction by 2. A further 
reduction by 2 occurs due to the conventional Parisi-Sourlas mechanism.

Recently, the problem of rooted spiral site trees embedded in a triangular
lattice has been studied by exact enumeration\cite{Santra2}. It will be of
interest to verify whether the dimensional reduction by 4 occurs in this case
also. The universality of the exponents $\theta$ and $\nu$ imply that they
have the same values as long as the dimension d of the underlying lattice
is the same. Also, the values remain unchanged if one considers bond LAs
instead of site LAs. These facts are well-established for both the undirected
and directed LA problems. Preliminary results in the case of the spiral LA
problem, however, indicate that the values of $\theta$ and $\nu$ are
different for the square and triangular lattices. If the result is proved to
be true, then one has an example of the breakdown of universality in the
case of the LA problem. 

We next consider the configurational statistics of spiral site LAs with c
(c = 0,1,2,...) number of loops. The corresponding growth parameter and the
exponents are $\lambda_c$, $\theta_c$ and $\nu_c$ respectively.The number
of loops c in an animal of s sites has been determined through the relations:
\begin{equation}
b\,=\,1/2\,\sum_{i=1}^{z}\,i\,q(i)\,\,\,\,,c\,=b\,-\,s\,+\,1
\end{equation}
where b is the number of bonds in the animal, z is the coordination number
of the lattice and q(i) is the number of sites in the animal which have i
nearest-neighbour occupied sites. In the case of undirected and directed
LAs, $\lambda_c$ and $\nu_c$ are independent of c and equal to $\lambda_0$
and $\nu_0$, the respective values for trees. The animal number exponent 
$\theta_c$ is related to that for trees ($\theta_0$) through the relation
\cite{Duarte,Whittington,Lam}
\begin{equation}
\theta_c\,=\,\theta_0\,-\,c
\end{equation}
A rigorous proof of (18) has been given for undirected weakly-embeddable
animals ( bond animals) in the square lattice\cite{Soteros}. Universality
demands that the same relation be true for strongly-embeddable (site)
animals. For spiral LAs, no rigorous result is available but calculations
based on exact enumeration data suggest the relation
\begin{equation}
\theta_c\,=\,\theta_0\,\,-\,\alpha\,c
\end{equation}
where $\alpha$ = 1.5 $\pm\,0.2$.Whittington et al \cite{Whittington} have
derived the inequality
\begin{equation}
\theta_c\,\geq\,\theta_{c+1}\,\geq\,\theta_c\,-1
\end{equation}
The inequality is derived from another inequality involving LA numbers
\begin{equation}
g_{s-3\,,c}\,\leq\,g_{s\,,c+1}\,\leq\,2dsg_{s,c}
\end{equation}
where $g_{s,c}\,\sim\,\lambda_c^s\,s^{-\theta_c}$. For
spiral site animals, an inequality similar to the first half of the inequality
in (21) can be proved\cite{Santra1}.Let the vertices or sites of the spiral
animals have coordinates $(x_i,y_i)$, i=1,2,....,n.Following Whittington
et al\cite{Whittington}, the top (bottom) vertex is defined as the vertex 
having maximum (minimum) x coordinate and, in case of ambiguity, the vertex 
in this subset having maximum (minimum) y coordinate. Consider any c-loop
animal belonging to the set $W_{s,c}$ of animals with s sites and c loops. The
number of animals is $g_{s,c}$. Let t be the top vertex of the animal with
coordinates $(x_t,y_t)$. The top vertex can be approached in three ways: 
(i) from the vertex ($x_t\,-\,1,\,y_t$) if it is occupied,
(ii) from the vertex $(x_t\,,\,y_t\,-1)$ if it is occupied,
(iii) from both the vertices ($x_t\,-1\,,y_t$) and $(x_t\,,\,y_t-1)$ if
both are occupied.

In cases (ii) and (iii), add four vertices $v_1,v_2,v_3,v_4$ with 
coordinates $(x_t\,+\,1,y_t),\,(x_t\,+\,1,\,y_t\,-1),\,(x_t\,+\,2,\,y_t)$
and $(x_t\,+\,3,\,y_t)$ respectively. For lattice site animals, addition
of a new site implies that the bonds between this site and its nearest
occupied neighbours automatically exist. Addition of four new sites leads
to an increase in the number of bonds by five. The number of loops 
therefore increases by one. The new animal obtained is also spirally
connected and so belongs to the set $W_{s+4,c+1}$ of animals with c+1
loops. In case (i), there are two possibilities, either the vertex
$(x_t\,,\,y_t-1)$ is occupied or unoccupied. If the first possibility
is true, add four vertices with the same coordinates as in the cases of
(ii) and (iii) to get a spiral animal of s + 4 sites and c+1 loops. If 
the second possibility is true, add vertices with coordinates $(x_t+1,y_t),
(x_t\,+\,2,\,y_t),\,(x_t\,+\,1\,,\,y_t-1)$ and $(x_t\,+\,2,\,y_t-\,1)$
generating a spiral animal of s + 4 sites and c + 1 loops. Thus for each
animal belonging to the set $W_{s,c}$ of animals, an animal belonging to the set
$W_{s+4,c+1}$ can be generated ,i.e.,
$g_{s,c}\,\leq\,g_{s+4,c+1}$ or, $ g_{s-4,c}\,\leq\,g_{s,c+1}$.      

The proof given by Whittington et al\cite{Whittington} for the second
half of the inequality fails in the case of spiral animals. For undirected
weakly embeddable animals, an upper limit to $g_{s,c}$ is $2dsg_{s,c-1}$ as
there are s vertices and at each vertex an edge can be added in 2d ways,
addition of an edge (bond) implying an increase in the number of loops by
1. Thus $g_{s,c}\,\leq\,2dsg_{s,c-1}$, or, $g_{s,c+1}\,\leq\,2dsg_{s,c}$, the
second half of the inequality in (21), is obtained. In the case of spiral
lattice animals, animals with c loops can be generated not only from spiral
animals with c-1 loops but also from undirected animals with c - 1 loops, which on the 
addition of an edge( bond) become spirally connected with c loops. Thus
a simple relation like $g_{s,c}\,\leq\,2dsg_{s,c-1}$ cannot be written down for
spiral lattice animals.

For spiral LAs, no rigorous results are available. This is in contrast to the
case of directed LAs. In this context, we mention a directed compact site LA 
problem on the d-dimensional hypercubic lattice\cite{Wu}. It has been 
rigorously established that the model is equivalent to the (i) infinite-state
Potts model and (ii) the enumeration of (d-1)-dimensional restricted partitions
of an integer. Thus, the directed compact LA problem can be solved exactly
in d = 2,3 using known solutions of the enumeration problem mentioned. Also, 
knowledge of the infinite-state Potts model solution leads to a conjectured
limiting form, for d >3, of the generating function of restricted partitions.
The latter problem is still unsolved in number theory. The examples of SSAW
and the directed compact site LA problem show that the well-known 
mathematical theory of partition of integers can be used to derive rigorous
results for lattice-statistical models.

\section*{IV. Spiral site percolation}

Santra et al\cite{Santra3,Santra4} have formulated a new type of percolation 
model known as spiral percolation in which a rotational constraint is
operative due to which each step in a percolation path is either in the
forward direction or in a specific rotational direction, say, clockwise.
Consider spiral site percolation on a lattice in 2d of size L$\times$L.
The spiral percolation threshold is determined by the binary search method
\cite{Stauffer1}. One starts from a central site of the lattice called
the origin. Initially, from the origin, one can proceed in any one of the four 
possible directions. The nearest-neighbour (n.n.) sites in these directions
are occupied with probability p by using a random number generator. With
each occupied site a variable JVISIT(IV,INOW) is associated. INOW is the
site index and IV the direction index. The sites of the lattice are numbered 
according to a particular sequence; the site index is the number associated
with a site. A site can be reached from four directions,south,west,north and 
east. The corresponding direction indices are 1,2,3 and 4 (Figure 5(a)).
The rotational constraint implies the following: if a site has direction
index 1, then occupy with probability p, only those of the n.n.s which are 
towards the north and east of the site. Initially, the JVISIT variable is
assigned the value zero for all sites. As soon as a site is occupied, the 
corresponding JVISIT variable is given the value IV where IV is the direction
index corresponding to the direction from which the site is occupied. All
growing sites are put on a list and the walk-search is carried out for each.
The growth of a cluster of occupied sites stops only when all the perimeter
sites are unavailable for occupation. Figure 5(b) shows a typical cluster 
grown obeying the rotational constraint. Due to the nature of the constraint,
loops are an essential feature of the growing spirally connected cluster.
In the computer algorithm, the possibility of loop formation while growing a
cluster is taken care of in the following manner. Refer to Figure 5(b).The
site i is the origin. In the first step when j is occupied, JVISIT(i,j) =1. Next time
j is approached (from site m,say), the computer algorithm checks whether
JVISIT(4,j) is zero. If it is non-zero then the site j has been approached
from the east previously and so cannot be occupied again. For the particular
cluster in Figure 5(b), since JVISIT(4,j) is zero, the site j is
reoccupied and JVISIT(4,j) assigned the value 4. Thus loop formation and
continuation of cluster growth are possible. 

Following the binary search method, one starts growing a cluster with a 
particular value of p,say $p_0$. If the cluster grown spans (does not span)
the lattice in either the east-west or north-south direction, $p_0$ is
decreased (increased) by a small amount. The same random number sequence
is then used to get estimates $p_1(L)$ and $p_2(L)$ which bound an interval
containing the true threshold value $p_c(L)$. By successive binary chopping
of this interval one can determine $p_c(L)$ with a specified accuracy. The
whole process is then repeated N times ($N\,\times L\times L$ is of the
order $(10^6\,-\,10^7)$) using different random number sequences. The average
value $<p_c(L)>$ of all the estimates obtained is taken as an estimate for
the percolation threshold. 

We now describe the scaling behaviour of the percolation clusters in the
vicinity of the percolation threshold. In the case of undirected percolation,
the most widely used method of calculating the scaling behaviour is to fill up 
a large lattice by clusters of various sizes s with site occupation 
probability p\cite{Stauffer1}. The cluster size distribution is given by $n_s(p)$,
the number of clusters of size s per site of the lattice.The probabilty
that a site belongs to a cluster of size s is $sn_s(p)$. The probability
that an arbitrary site belongs to any cluster is equal to the probability
that it is occupied. Thus
\begin{eqnarray}
\sum_{s}\,s\,n_s(p)\,&=&\,p\,\,(p\,<\,p_c)\\
\sum_{s}\,s\,n_s(p)\,+\,P_{\infty}(p)\,&=&\,p\,\,(p\,>p_c)
\end{eqnarray}
The sums in (22) and (23) run over all possible finite cluster sizes.$P_{\infty}(p)$
is the probability that an arbitrary site (occupied or not) belongs to the 
infinite cluster. The percolation probability P(p) is defined as the probability
that an occupied site belongs to the infinite cluster, i.e., 
\begin{equation}
P(p)\,=\,P_{\infty}\,/p\,=\,1\,-\,\sum_s\,s\,n_s(p)
\end{equation}
Let W = $(sn_s)/\,(\sum_s\,sn_s\,)$ be the probability that the cluster to
which an arbitrary occupied site belongs contains exacly s sites. The average
cluster size $S_{av}$ is then given by
\begin{equation}
S_{av}\,=\,\sum_s\,s\,W_s\,=\,\sum_s\,(S^2n_s)/(\sum\,sn_s)
\end{equation}
The linear size of the finite clusters, below and above $p_c$, is characterised
by the correlation length $\xi$. The correlation length is defined as the
mean distance between two sites on the same finite cluster. As the site
occupation probability p reaches the percolation threshold $p_c$, the various moments of the cluster size
distribution $n_s(p)$  given by
\begin{equation}
\sum^{\prime}\,s^k\,n_s(p),\,\,\,\,\,k\,=\,0,1,2,...
\end{equation}
(the prime denotes that the largest cluster is excluded from the sum) become
singular with characteristic critical exponents. The critical exponents
$\alpha,\beta,\gamma,\delta,\nu$ are defined through the relations
\begin{equation}
k\,=\,0:\,\,\,\,(\sum_s^{\prime}\,n_s)_{sing}\,\sim\,\mid\,p\,-\,p_c\mid^{2-\alpha}
\end{equation}
\begin{equation}
k\,=\,1:\,\,\,\,P(p)\,\propto\,(\sum_s^{\prime}sn_s)_{sing}\,\sim\,(p-p_c)^\beta
\end{equation}
\begin{equation}
k\,=\,2:\,\,\,S_{av}\,\propto\,\sum_s^{\prime}\,s^2\,n_s\,\sim\,\mid p-p_c\mid^{-\gamma}
\end{equation}
\begin{equation}
\xi\,=\,\mid\,p\,-p_c\,\mid^{-\nu}
\end{equation}
\begin{equation}
\sum\,sn_s(p_c)\,(\,1\,-e^{-hs})\,\sim\,h^{1/\delta} 
\end{equation}
The subscript `sing' refers to the leading non-analytic contribution. The
above equations are valid in the limit in which the fictitious field h 
and the quantity $\mid\,p\,-p_c\,\mid$ both go to zero.                                              

We now consider the case of spiral site percolation. The scaling behaviour
of clusters in the critical region is studied following the single-cluster 
growth method in which clusters are grown singly starting from a fixed origin
instead of filling up the whole lattice by clusters of various sizes. The
computer algorithm has already been described. The cluster size distribution
is now defined as
\begin{equation}
P_s\,(p)\,=\,N_s\,/\,N_{tot}
\end{equation}
where $N_s$ is the number of clusters of size s in a total number $N_{tot}$
of clusters generated. The various moments of $P_s(p)$, 
$\sum_s^{\prime}\,s^k\,P_s(p)$ becomes singular as $p\rightarrow p_c$. The
prime in the summation symbol indicates that the largest cluster is to be
left out of the sum. The average cluster size $\chi$ is given by
\begin{equation}
\chi\,\sim\,\sum_s^{\prime}\,s\,P_s(p)
\end{equation}
which diverges as $\chi\,\sim\,\mid\,p\,-\,p_c\mid^{-\gamma}$ as
$p\,\rightarrow p_c$. In the single cluster growth method, the cluster
configurations are rooted at the origin and the first, not the second,
moment of the cluster size distribution defines the exponent $\gamma$.
In fact, any exponent which corresponds to the kth moment of the cluster
size distribution in the case of the lattice-filling method, is obtained
from the (k-1)th moment, one order less, in the case of the single cluster
growth method. In the case of directed and spiral percolation, where
there is a directionality constraint on the percolation process, the single 
cluster growth method is the appropriate one for the study of critical
behaviour.

The percolation transition is a second-order phase transition and exhibits
the critical phenomena characteristic of second-order thermodynamic phase
transitions like liquid-gas and paramagnet-ferromagnet transitions. The
major characteristic feature, which we have already discussed, is that different 
cluster-related quantities exhibit power-law behaviour as $p\rightarrow p_c$.
One of these, the percolation probability P(p) (Eq.(24)) serves as the order
parameter of the transition. The order parameter has a non-zero value in one
phase (p > $p_c$) and is zero in the other phase (p < $p_c$). As in the case
of thermodynamic phase transitions, the critical behaviour is universal, i.e.,
the magnitude of the critical exponents depends only on the dimensionality 
of the lattice and is the same for both site and bond percolation. The critical
exponents are not independent (only two are so) and one can write down `scaling'
relations connecting different critical exponents. The scaling relations can be
determined on the basis of the scaling hypothesis according to which in the vicinity
of p = $p_c$, the cluster-related quantities become generalised homogeneous
functions (GHF). A function F(x,y) of two variables is a GHF if it satisfies 
the property
\begin{equation}
F(\,x\,,y)\,=\,x^A\,f(y/x^B)
\end{equation}
where the function f is the appropriate scaling function and is the function
of a single variable. The form (34) usually holds true when both the
variables x and y tend to zero. The definition (34) of a GHF can be 
generalised to more than two variables. A second-order thermodynamic phase
transition has the feature that fluctuations occur at all length scales, i.e.,
the correlation length goes to infinity. In the case of percolation
transition also, the correlation length goes to infinity. The infinite
percolation cluster is a self-similar object with fractal dimension less than 
the Euclidean dimension of the space in which the cluster is embedded.

We now describe the approximate methods for calculating the critical exponents
and determining the scaling behaviour of the cluster-size disribution function.
The methods normally used are Monte Carlo (MC) simulation, finite-size scaling,
series-expansion and renormalization group. We consider the case of spiral site
percolation on the square and triangular lattices and describe the results
obtained by Santra and Bose\cite{Santra3,Santra4}. In MC simulation, a 
particular lattice size is considered. The value of the percolation threshold 
$p_c$ is determined by the binary search method. Ten thousand cluster are generated
by the single cluster growth method. The growth of a cluster is stopped when
there is no further site available for occupation. The quantities, the
average cluster size $\chi$ (Eq.33 ), the second moment $\chi^{\prime}$ of
the cluster size distribution $P_s(p)$  and the correlation length
$\xi$ are determined as a function of $\mid p - p_c\mid$. As the site 
occupation probability p tends towards the percolation threshold, $\chi$,
$\chi^{\prime}$ and $\xi$ diverge with the exponents $\gamma,\gamma^{\prime}$
and $\nu$. We first quote the results for the square lattice. The value of
$p_c$ is $p_c$ = 0.711 $\pm$ 0.001. The critical exponent $\gamma$ is 
obtained from the slope of $log\chi$ versus $log\mid\,p\,-p_c\mid$ and is
given by $\gamma$ = 2.19 $\pm$ 0.07.The exponents $\gamma^{\prime}$ and
$\nu$ are $\gamma^{\prime}$ = 4.51 $\pm$ 0.16 and $\nu$ = 1.16 $\pm$ 0.01.
The exponent $\nu$ can also be determined from the scaling relation
$\nu\,=\,(\gamma^{\prime}\,-\,\gamma)\,/2$ as  $\nu$ = 1.18 $\pm$ 0.23. This
value is in agreement with the value obtained through direct measurement.
For the triangular lattice, $p_c\,=\,0.667\,\pm\,0.001,\,\gamma\,=\,
2.079\,\pm\,0.062,\,\,\gamma^{\prime}\,=\,4.376\,\pm\,0.199 $ and
$\nu\,=\,1.012\,\pm\,0.025$.

True critical behaviour occurs only in the limit of infinitely large
lattices. An estimate of the critical exponents can, however, be obtained
from the studies of finite systems by assumung the finite-size scaling
hypothesis\cite{Stauffer1} which leads to the formula
\begin{equation}
A\,=\, L^{-x/\nu}\,\,F\,[\,(p-p_c)\,L^{1/\nu}]
\end{equation}
where A is a quantity which becomes critical, A $\sim\,\mid\,p\,-\,p_c\mid^x$
as $p\,\rightarrow\,p_c$. The function F is a suitable scaling function.
At p = $p_c$, the quantity A varies as $L^{-x/\nu}$. This result can be used
to determine the exponents $\gamma/\nu,\,\gamma^{\prime}/\nu$ by calculating
the first and second moments $\chi$ and $\chi^{\prime}$ respectively, of the
cluster size distribution. The percolation threshold $p_c$ = <$p_c(L)$>, the
average being taken over a large number of estimates of $p_c$ for a $L\times L$
lattice. The largest cluster spanning the lattice is of size $S_L$ and is a
fractal with fractal dimension $d_f$ defined by
\begin{equation}
S_L\,\,\sim L^{d_f}
\end{equation}
The fractal dimension $d_f$ can again be written as
\begin{equation}
d_f\,=\,d\,-\,\beta/\nu
\end{equation}
where $\beta$ is the exponent associated with the order parameter. From (36), the fractal dimension
$d_f$ can be determined and from (37) putting d = 2, the value of $\beta/\nu$
can be calculated. The measurement of various quantities is done for lattices of
various sizes. The slope of the straight line log$S_L$ versus logL gives
$d_f\,=\,1.965\,\pm\,0.009$ and so $\beta/\nu\,=\,0.043\,\pm\,0.009$. From
(35), the plot of logA versus logL at p = $p_c$ gives the exponent $x/\nu$.
Thus for A = $\chi$, the average cluster size, the exponent $\gamma/\nu$ = 
2.01 $\pm$ 0.06. For A = $\chi^{\prime}$, the second moment of the cluster
size distribution, the exponent $\gamma^{\prime}/\nu$ = 4.05 $\pm$ 0.13.
For the triangular lattice, the various quantities have been determined as
$d_f\,=\,1.965\,\pm\,0.008,\,\beta/\nu\,=\,0.035\,\pm\,0.008,\,\gamma/\nu
\,=\,1.867\,\pm\,0.028$ and $\gamma^{\prime}/\nu\,=\,3.829\,\pm\,0.062$.

We now describe the method of series expansion. The probability p that the origin
from which clusters grow is occupied can be written as a sum over all
finite clusters that start from it (for p < $p_c$), i.e.,
\begin{equation}
p\,=\,\sum_{s,t}\,g_{st}\,p^s\,(1-p)^t\,=\,\sum_s\,p^s\,D_s(q),\,\,q\,=\,1-p
\end{equation}
where $g_{st}$ is the number of clusters or LAs of s sites and t perimeter
sites and $D_s(q)$'s are the perimeter polynomials. The animals are rooted at
the origin. The average cluster size $\chi$ is the first moment of the cluster
size distribution.
\begin{equation}
\chi\,=\,\sum_{s,t}\,s\,g_{st}\,p^s\,(1-p)^t\,\sim\,\mid p\,-\,p_c\mid^{-\gamma}
\end{equation}
as p $\rightarrow\,p_c$. The clusters rooted at the origin obey the same
rotational constraint as in the percolation process. The spiral LAs, as
already mentioned in Section III, have been extensively studied by Bose et al
\cite{Bose1,Bose2,Santra1}. The major focus in these studies has been on 
spiral site LAs with no fixed origin. If the origin is not kept fixed, the
number of perimeter sites is dependent on the position of the origin. In
spiral percolation, the percolation clusters are rooted at the origin so the spiral
LAs in the series-expansion method have to be rooted, i.e., have a fixed
origin. Now each distinct LA configuration has a unique value for the number of
perimeter sites. The rooted spiral site LAs on the square lattice have been 
enumerated exactly by Santra and Bose\cite{Santra4} for size upto 13. Thus, 
series expansion for the average cluster size $\chi$ (eq.(39)) is exactly
known, when expanded in powers of p, upto the term containing the power
$p^{13}$. A standard trick \cite{Santra4} enables one to extend the series by
one more power. Thus the coefficients in the expansion of $\chi\,=\,1\,+\,
\sum_r\,b_r\,p^r$ (a common factor of p has been taken out), are exactly
known for r = 1 to 13. As $p\,\rightarrow\,p_c$, $\chi$ has the power-law
behaviour $\chi\,\sim\,(p_c\,-\,p)^{-\gamma}\,=\,p_c^{-\gamma}\,(1-p/p_c)^{-\gamma}$.
On binomial expansion, the ratio of coefficients $b_{k+1}/b_k\,=\,
1/p_c\,(1\,+\,(\gamma-1)/(k+1))$. The plot $b_{k+1}/b_k$ versus 1/(k+1)
should be a straight line. The intercept of the line on the y-axis gives
$1/p_c$ and from the slope, knowing $p_c$, the exponent $\gamma$ can be
determined. A more sophisticated method of obtaining the value of $\gamma$
is that of Pad\'{e} approximants. Using this method, the value of 
$\gamma$ for spiral site percolation on the square lattice has been obtained
as $\gamma\,=\,2.167\,\pm\,0.004$. For lack of a sufficiently long series,
the series-expansion method has not been applied to the triangular lattice.
The series-expansion method can be applied to generate the series for other
percolation-related quantities and thereby obtain the values of the 
associated exponents.

We now discuss the scaling hypothesis for percolation clusters. From Eq.(34),
one finds that if the function $G(z)\,=\,F(x,y)/\,x^A$ is plotted as a function
of z = $y/x^B$, all the data (for different values of x) fall on a single
curve described by the function f(z). On the other hand, if Eq.(34) were 
not true, the data, for different values of x, fall on separate curves. The
collapse of data on a single curve serves as a verification of the scaling 
hypothesis. The cluster size distribution function for sufficiently large
s has the scaling form in the critical region given by\cite{Santra4}
\begin{equation}
P_S(p)\,=\,s^{-\tau+1}\,f\,[(p-p_c)\,s^\sigma\,]
\end{equation}
where both p - $p_c$ and 1/s go to zero as $p\rightarrow\,p_c$. The exponents
$\tau$ and $\sigma$ are related to the exponents $\gamma$ and $\beta$
through the relations
\begin{equation}
\gamma\,=\,(3-\tau)/\sigma\,\,\,\,,\beta\,=\,(\tau-2)/\sigma
\end{equation}
A verification of the scaling function form (40) is possible by plotting
$P_s(p)/P_s(p_c)$ against $ (p-p_c)s^\sigma$. If the scaling form is true
then, for sufficiently large clusters and for different values of p, the
data should collapse onto a single curve. For the square and triangular
lattices, the scaling form (Eq.(40)) for spiral site percolation has
been verified with $\sigma\,=\,0.446$ (square lattice) and $\sigma\,=\,
0.473$ (triangular lattice).

The values of the critical exponents obtained for spiral site percolation
in 2d are different from the values of critical exponents in the case of 
undirected percolation. Spiral percolation is similar in nature to directed
percolation, since for both the models a directional constraint is operative.
In the first case, the directional constraint is rotational in nature 
and in the second case percolation occurs only in certain specific 
directions. Dhar and Barma\cite{Dhar2} have studied directed site percolation 
on the square lattice using MC simulation. A comparison with their results
shows that the average cluster size exponent $\gamma\,=\,2.19\,\pm0.03$ 
(directed percolation) is the same as that obtained in the case of spiral
percolation, $\gamma\,=\,2.19\,\pm0.07$, using MC simulation. Thus the 
average cluster size diverges with the same exponent irrespective of the
nature of the external constraint.One crucial difference between spiral
and directed percolation is that in the first case. percolation clusters grow 
isotropically whereas in the second case, the clusters grown are 
anisotropic in nature. For the first case, there is thus only one correlation
length which diverges with the exponent $\nu$ as $p\rightarrow p_c$, 
whereas, in the second case, there are two correlation lengths, one parallel
and the other perpendicular to the preferred direction, which diverge
with the exponents $\nu_{\parallel}$ and $\nu_{\perp}$ respectively as
$p\rightarrow\,p_c$.

One important quantity in the percolation problem is the cluster external
perimeter or `hull' of the large clusters. The hull of a cluster is the
continuous path of occupied sites at the external boundary of the cluster.
The hull has a significance of its own apart from being a part of the 
percolation cluster.Sapoval et al\cite{Sapoval} have shown that the 
diffusion front arising out of diffusion from a source has a fractal structure
that is related to the hull of percolation clusters. The percolation cluster
hull in the case of undirected percolation exhibits scaling behaviour 
characterised by critical exponents which have values different from those of 
analogous quantities in the case of percolation clusters. Santra and Bose
\cite{Santra5} have studied the scaling behaviour of the percolation hull
of large clusters in the case of spiral site percolation on the square and
triangular lattices. The scaling behaviour is found to be different from that 
of hull in the case of undirected percolation. In the latter case, Saleur
and Duplantier\cite{Saleur} have proved that the fractal dimension, $d_H$,
of the hull is given by
\begin{equation}
d_h\,=\,1\,+\,1/\nu
\end{equation}
where $\nu$ is the correlation length exponent. Eq.(42) is an exact result.
A similar relation cannot be proved in the case of spiral site percolation.

\section*{V.Concluding Remarks}

The rotational constraint, we have seen, has a non-trivial effect on the 
statistics of lattice models like SAW, LA and percolation. The undirected
SAW and LA problems have been extensively studied and some fairly
rigorous results are available. For undirected percolation in 2d, the
values of $p_c$ are known exactly in some cases. The critical exponents are,
through some exact mappings, conjectured to be known exactly\cite{Isichenko}.
For the directed SAW,LA and percolation problems, many exact/analytical results are 
available. In the presence of a rotational constraint, however, the only
rigorous results that are known are for the SSAW model. The spiral LA and
spiral percolation problems have been studied through numerical methods.
Derivation of numerical results for the models is a challenge for both
mathematicians and physicists. The computational methods in the case of
spiral models are also more difficult to implement than in the case of
undirected and directed models.

Lattice models have extensive applications in all branches of science.The
SAW model describes a correlated walk, i.e., a walk with memory. The
knowledge gained from the study of the model has been found to be useful
in problems like correlations in nucleotide sequences\cite{Goldberger} and
the folding of proteins\cite{Chan}. Several examples of the various 
applications of the lattice models can be found in the Refs.[ 1-5].Examples
of the spiral structures and the rotational constraint can be found in Ref.
[41]. A few examples are: spiral galaxies, the pattern of
shoot arrangement in plants, scales on a pineapple or on a pine cone, the 
Belousov-Zhabotinsky reaction, plasma in a stochastic magnetic field, 
polymers with spiral structure, bacterial colonies which exhibit chiral
morphology\cite{Ben-Jacob} (the colony consists of twisted branches all
with the same handedness) etc. It is to be hoped that further studies
will be undertaken to understand the effect of rotational constraint on
the properties of model systems.

{\large Acknowledgement}

I thank Asimkumar Ghosh for help in preparing the manuscript.

\newpage
\section*{Figure Captions}
\begin{description}
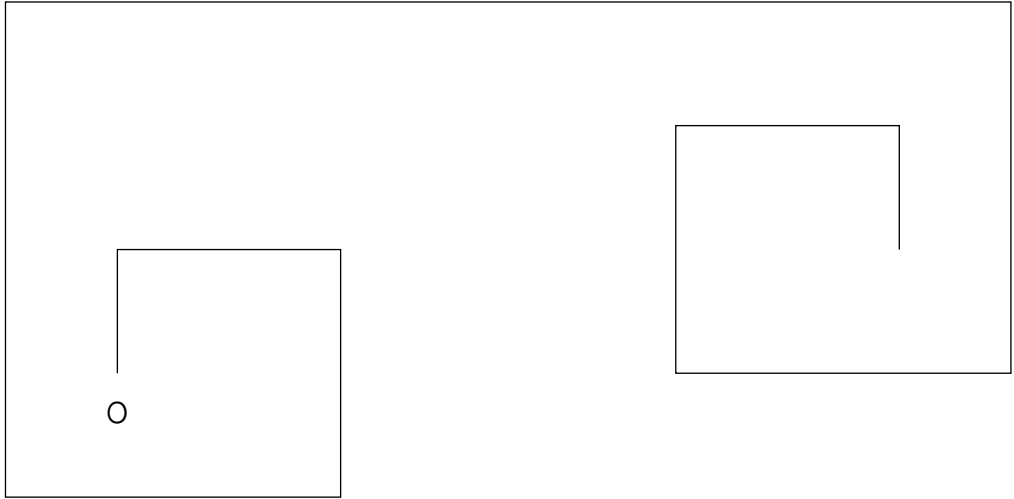
\item[Fig.1] A spiral self-avoiding walk with an outward spiralling part
and an inward spiralling part; `O' denotes the origin.
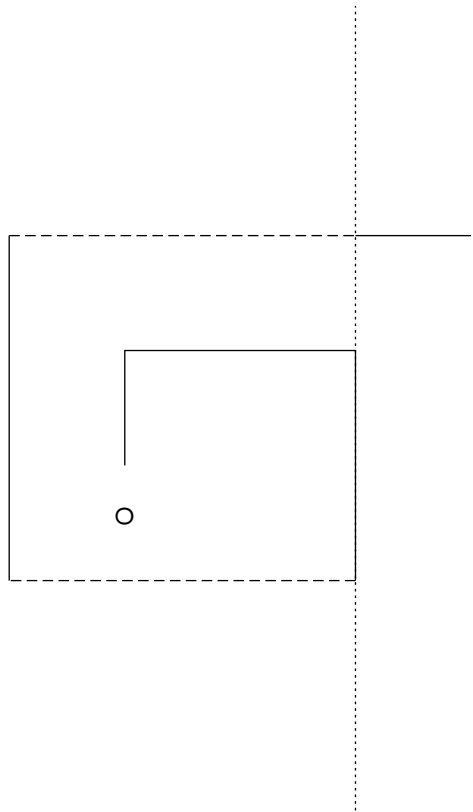
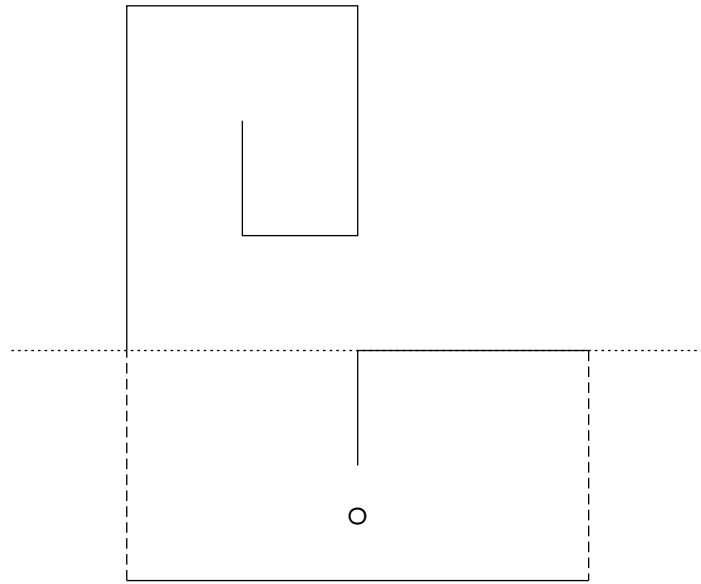
\item[Fig.2] Three different spiral self-avoiding walks with origin O and
end-point N. The broken line divides the walk into two parts.
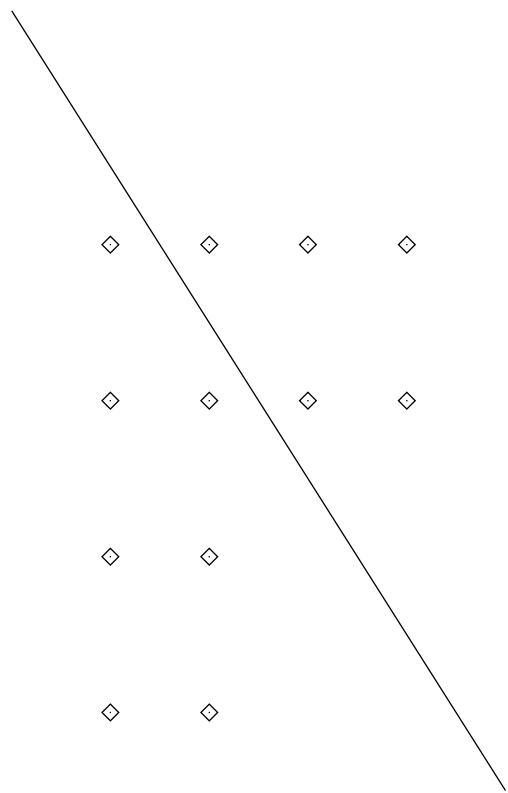
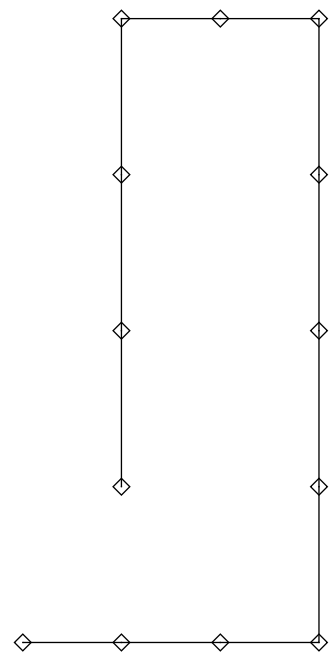
\item[Fig.3](a) A partition of the number 12 into the integers 4,4,2 and 2.
(b) the spiral self-avoiding walk corresponding to the set of points in (a).
\item[Fig.4] A spiral lattice animal of 10 sites on a square lattice grown 
from the origin X.
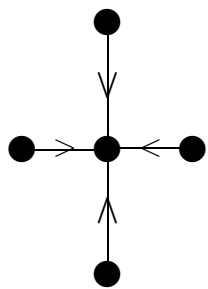
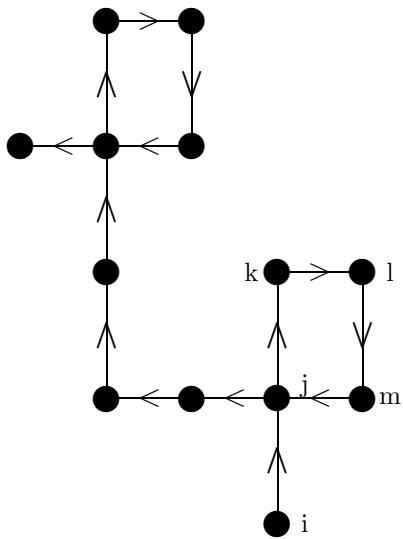
\item[Fig.5](a) Direction indices of a site corresponding to the four
different directions from which the site can be reached, (b) an example
of a spiral site cluster on the square lattice. The arrows on the bonds
indicate the allowed spiral directions of flow from site i.
\end{description}
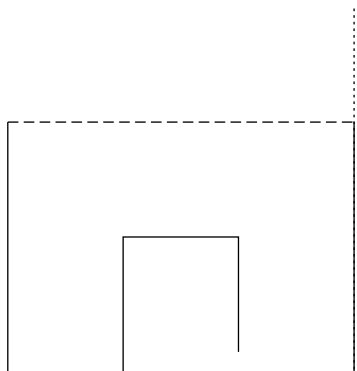
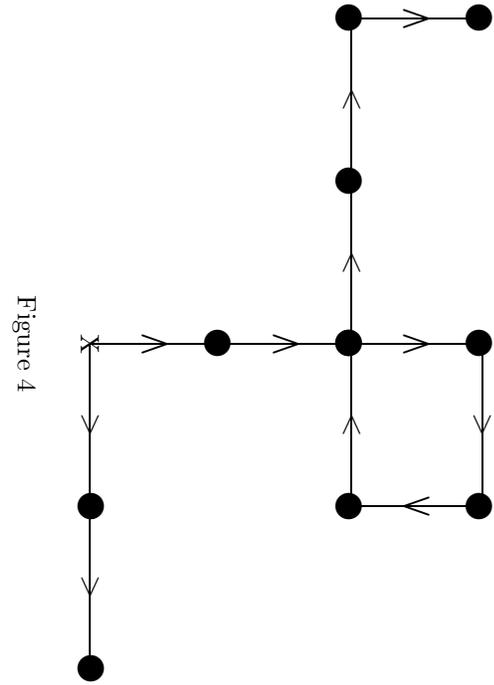
\end{document}